\documentclass[aps,prl,twocolumn,showpacs,preprintnumbers,superscriptaddress,floatfix,amsmath,amssymb,amsfonts,10pt,footinbib,longbibliography]{revtex4-1}
\usepackage[utf8]{inputenc}
\usepackage{amsmath,amsfonts,amssymb}
\usepackage{graphicx}
\usepackage{color}
\usepackage[sf]{subfigure}
\usepackage{fouriernc}
\usepackage{xcolor}
\usepackage{bm}

\definecolor{linkcolor}{RGB}{6,69,173}

\usepackage[colorlinks=true,
            linkcolor=linkcolor,
            urlcolor=linkcolor,
            citecolor=linkcolor,
            unicode,
            pdfencoding=auto]{hyperref}
\usepackage{etoolbox}

\begin{document}

\title{Pomeranchuk Instability of Composite Fermi Liquid}
\author{Kyungmin Lee}
\affiliation{Department of Physics, The Ohio State University, Columbus, Ohio 43210, USA}
\author{Junping Shao}
\affiliation{Department of Physics, Binghamton University, Binghamton, New York 13902, USA}
\author{Eun-Ah Kim}
\affiliation{Department of Physics, Cornell University, Ithaca, New York 14853, USA}
\author{F. D. M. Haldane}
\affiliation{Department of Physics, Princeton University, Princeton, New Jersey, USA}
\author{Edward H. Rezayi}
\affiliation{Department of Physics, California State University Los Angeles, Los Angeles, California 90032, USA}

\begin{abstract}
Nematicity in quantum Hall systems has been experimentally well established at excited Landau levels.
The mechanism of the symmetry breaking, however, is still unknown.
Pomeranchuk instability of Fermi liquid parameter $F_{\ell} \le -1$ in the angular momentum $\ell=2$ channel has been argued to be the relevant mechanism, yet there are no definitive theoretical proofs.
Here we calculate, using the variational Monte Carlo technique, Fermi liquid parameters $F_\ell$ of the composite fermion Fermi liquid with a finite layer width.
We consider $F_{\ell}$ in different Landau levels $n=0,1,2$ as a function of layer width  parameter $\eta$.
We find that unlike the lowest Landau level, which shows no sign of Pomeranchuk instability, higher Landau levels show nematic instability below critical values of $\eta$.
Furthermore, the critical value $\eta_c$ is higher for the $n=2$ Landau level, which is consistent with observation of nematic order in ambient conditions only in the $n=2$ Landau levels.
The picture emerging from our work is that approaching the true 2D limit brings half-filled higher Landau-level systems to the brink of nematic Pomeranchuk instability.
\end{abstract}

\maketitle

The electronic nematic order first conjectured in the context of doped Mott insulators \cite{kivelon-n-1998} has become a common electronic phase in the field of strongly correlated quantum matter as more of these  systems are found to exhibit the nematic order.
Electronic nematic ordering refers to a spontaneous breaking of spatial rotational symmetry while preserving translational symmetry.
Nematic ordered systems exhibit preferential direction, and ordering is often detected through anisotropy in longitudinal transport \cite{fradkin-arcmp-2010}.
The systems that exhibit nematic order now include underdoped cuprates, Sr$_3$Ru$_2$O$_7$, half-filled higher Landau-level states \cite{fradkin-arcmp-2010}, Fe-based superconductors \cite{fernandes-np-2014}, and even the surface of bismuth \cite{feldman-s-2016}.
On the one hand, such a ubiquity implies that the electronic nematic order fits into an over-arching classification of how strongly correlated electrons organize themselves.
In particular, the ubiquity underscores the original rationale for electronic liquid crystal phases based on the observation of frustration between kinetic energy and interaction energy and also, by  analogy, to the classical liquid crystalline systems.
On the other hand, this ubiquity motivates one to seek a microscopic mechanism of just how the analogy is realized.

\begin{figure}[t]
    \centering
    \subfigure[\label{fig:pomernem}]{\includegraphics[height=1.5in]{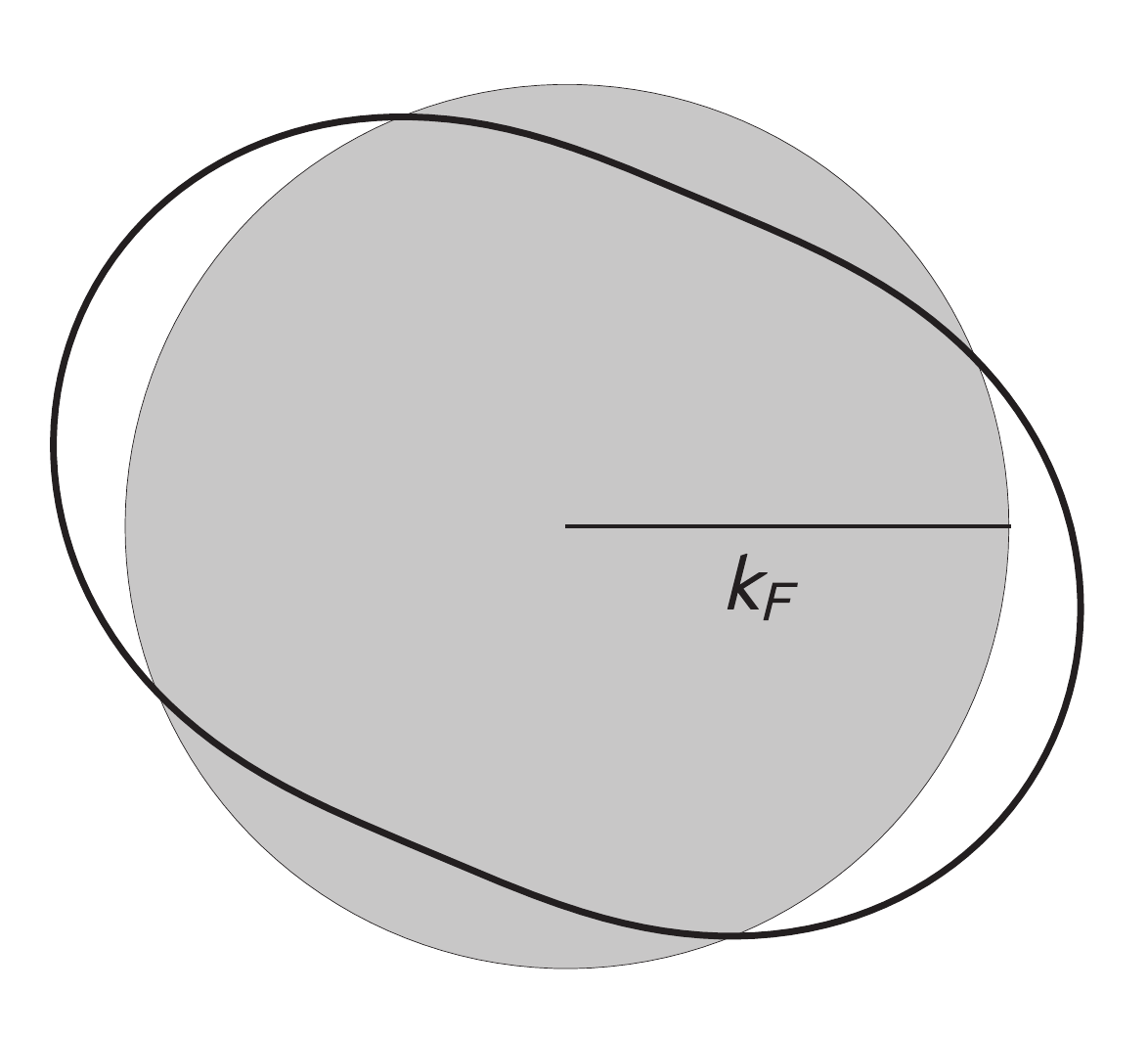}}
    \subfigure[\label{fig:excon}]{\includegraphics[height=1.5in]{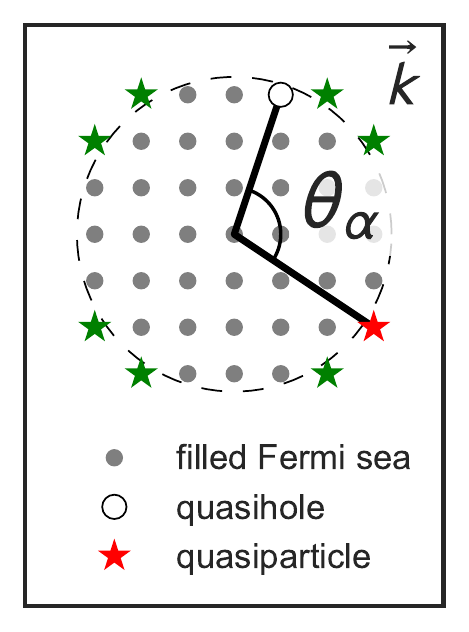}}
    \caption{\label{fig:1}%
(a) Deformation of the Fermi surface in the angular momentum $\ell=2$ (nematic) channel.
(b) Filled Fermi sea of composite fermions and a quasiparticle-quasihole pair configuration with the lowest energy marked in red. 7 other configurations with the quasihole momentum and the same kinetic energy are marked in green.
}
\end{figure}

Although the original picture of a nematic order forming through quantum melting \cite{kivelon-n-1998} (or impurity driven inhibition \cite{nie-pnas-2014}) of a stripe order is intuitively appealing, it has been difficult to make theoretical progress from this perspective (beyond phenomenology).
Instead, much progress in understanding the implications of nematic order relied on the notion of Pomeranchuk instability \cite{pomeranchuk-jetp-1958}.
Pomeranchuk pointed out that when a Fermi liquid parameter $F_{\ell}$ in the angular momentum $\ell$ channel for spin-polarized systems is less than -1, the Landau Fermi liquid is unstable against deformation of the Fermi surface in that channel.
Should microscopic interactions amount to $F_2<-1$, an isotropic Fermi liquid state would give way to a nematic state with an elliptically deformed Fermi surface [Fig.~\ref{fig:pomernem}].
Unfortunately it is rather challenging to calculate Fermi liquid parameters from a microscopic Hamiltonian in strongly correlated systems.
Hence, past studies put in the  the value of $F_2=-1$ ``by hand'' as a guarantee for the nematic ground state \cite{oganesyan-prb-2001,yamase-prb-2005,khavkine-prb-2004,lawler-prb-2006,fischer-prb-2011}.

Here, we turn to the half-filled Landau levels (HFLLs)  where nematic order may border a non-Abelian quantum Hall state. 
Although the lowest Landau level remains featureless and gapless, at $\nu=5/2$ (the $n=1$ Landau level) a two-dimensional electron fluid under magnetic field shows a quantum Hall (QH) plateau that is widely believed to be associated with the non-Abelian 
Moore-Read (MR) QH state \cite{moore-npb-1991}.
Interestingly, application of an in-plane magnetic field \cite{pan-prl-1999,lilly-prl-1999b,friess-prl-2014,shi-prb-2015,xia-prl-2010,xia-np-2011,liu-prb-2013} or anisotropic strain \cite{koduvayur-prl-2011} closes the gap, leaving the system in an anisotropic fluid state.
Surprisingly, a recent experiment showed that even isotropic pressure can drive the $\nu=5/2$ QH state into a gapless state with anisotropic transport \cite{samkharadze-np-2016,schreiber-prb-2017}.
Moreover, at fillings $\nu=9/2$ and higher, gapless anisotropic transport \cite{lilly-prl-1999a,du-ssc-1999,pan-prb-2014} has been interpreted as evidence of electronic nematic order \cite{fradkin-prb-1999,fradkin-prl-2000}.

The observation of nematic phenomena motivated variational studies comparing energies of candidate states using
  the Hartree-Fock (HF) \cite{koulakov-prl-1996,*koulakov-prb-1996, *fogler-prb-1997, moessner-prb-1996}
  or variational Monte Carlo method \cite{doan-prb-2007},
  as well as exact diagonalization studies \cite{rezayi-prl-1999,rezayi-prl-2000}.
The Hartree-Fock calculations \cite{koulakov-prl-1996,*koulakov-prb-1996, *fogler-prb-1997, moessner-prb-1996} found the single-Slater determinant states with charge density wave order to have lower energies than the Laughlin-type liquid states for $n\geq 2$.
But then \textcite{doan-prb-2007} showed that anisotropically deformed composite fermion (CF) unprojected wavefunctions representing a nematic state have even lower energies for $n=2$, when the critical value of the layer ``thickness parameter'' $\eta$ is below a critical value.
Exact diagonalization studies in Refs.~\cite{rezayi-prl-1999,rezayi-prl-2000} showed that a ground state of up to 12 electrons in half-filled systems at $n\geq 2$ yields static structure factors that are strongly peaked at a finite wave vector that decreases with increasing $N$; such a gapless state gives way to the MR paired state \cite{moore-npb-1991} with infinitesimal additional pseudopotentials $V_1$ and $V_3$ for $n=1$.
Alternatively, there were efforts to investigate the implications of nematic quantum criticality using quantum field theory \cite{you-prx-2014, mesaros-prb-2016}.
Nevertheless, it has been unclear whether a simple screened Coulomb interaction potential can, in fact, drive Pomeranchuk instability spontaneously and whether higher Landau levels are susceptible to such an instability.
Here, we use a well-established many-body wave function for the CF Fermi liquid at half-filled Landau level to numerically evaluate Fermi liquid parameters $F_{\ell}$ for the lowest three half-filled Landau levels ($n=0,1,2$) using energy differences between various particle-hole pair excitation configurations [see Fig.~\ref{fig:excon}].
Thereby, we test the Pomeranchuk instability scenario for CFs under screened Coulomb interaction.

A wave function describing a filled Fermi sea \cite{halperin-prb-1993} of CFs \cite{jain-prl-1989} projected into the lowest Landau level was given (in the spherical geometry) by \textcite{rezayi-prl-1994}.
On a torus, the analogous state is given by:
\begin{align}
  |\Psi_{\text{CF}}(\{\bm k_i\})\rangle &=
    \det_{i,j} \left[ e^{i \bm k_i \cdot \bm R_{j}} \right]
    |\Psi_{1/2}\rangle,
\end{align}
where $|\Psi_{1/2}(\{\bm k_i\})\rangle$ is the bosonic Laughlin state at half filling \cite{haldane-bookchapter-1990} and $\bm R_{i}$ are the non-commutative guiding-center coordinates that act within a Landau level, independent of its index.
They satisfy the commutation rule $[R^a_i,R^b_j]= -i\epsilon^{ab}\ell_B^2$,where $\epsilon^{ab}$ is the antisymmetric symbol, and $\ell_B$ is the magnetic length.
The set of $\{ \bm k_{i} \}_{i=1, \ldots N_e}$, is single particle ``momenta'', where $N_e$ is the number of electrons in the system.
Periodic boundary conditions require that $\bm k$ satisfy $\exp( i \bm k \cdot \bm L_a )=1$, where $\bm L_a$ for $a=1,2$ are primitive translation vectors that specify the torus \cite{haldane-prl-1985}.
The set $\{ \mathbf{k}_{i} \}_{i=1, \ldots N_e}$ completely characterizes the many-body state with a total momentum $\mathbf{K}=\sum_i \mathbf{k}_{i}/Ne$ relative to the allowed values \cite{haldane-bookchapter-1990}.
The exponential factors in the determinant act as translation operators on $\lvert \Psi_{1/2}\rangle$ by displacing the $i$th particle by $ d^a = \epsilon^{ba} k_b\ell_B^2$.
It can be seen that under a  uniform boost of each $\mathbf{k}_i$ the above wave function remains invariant
(up to a phase and an overall  multiplicative constant).
This property is called $\bm K$ invariance \cite{haldane-unpub,shankar-prl-1997, halperin-prl-1998}.

The variational energy of the wave function is lowest if the set of $\{ \bm k_{i} \}$ are compactly clustered.
A phenomenological Hamiltonian that possesses clustering and $\bm K$ invariance properties was given by Haldane \cite{haldane-unpub},
\begin{equation}
  \mathcal H_0 =
\frac{\hbar^2}{2m^* N_e} \sum_{i<j} |\bm{k}_i-\bm{k}_j|^2,
\end{equation}
where $m^*$ is the effective mass of the composite fermions.
The Fermi liquid parameters of this model are all zero except $F_1=-1$.

The CF wave function $\Psi_{\text{CF}}$, however, is computationally prohibitive to use, particularly for Monte Carlo calculations because of its explicit antisymmetrization that requires $N_e!$ operations.
Therefore, we use an approximate wave function defined on a torus, which is analogous to the wave function in the spherical geometry by \textcite{jain-prb-1997}.
It was used by \textcite{shao-prl-2015} to calculate entanglement entropy.
For a system with $N_e$ electrons on a torus at half filling, the total flux through the system is $N_{\phi} = 2 N_e$.
The CF wave function in the symmetric gauge where the zeros of the Laughlin state are displaced by the  $\{ \bm d_i \}_{i=1 \ldots N_e} $ is then 
\begin{align}
F_{\mathrm{CF}}
    &=
    \det_{i,j}
        \left\{
            e^{-d_{j}^{*} z_{i}}
            \prod_{k(\neq i)}
            \sigma
            \left[
                z_{i} - z_{k} + 2(d_j - \bar{d})
            \right]
        \right\}
    \nonumber\\
&\quad
    \times
    F_{\mathrm{c.m.}}
    \left[ \sum_{i} ( z_{i} + \bar{d} ) \right]
    e^{-\sum_{i} z_{i} z_{i}^{*} / 2},
\end{align}
where
    $z_{i} \equiv (x_{i} + i y_{i})/\sqrt{2}\ell_B$,
    $\bm d_{i} \equiv (d_{i}^{x} + i d_{i}^{y})/\sqrt{2}\ell_B$, and
    $\bar{d} \equiv  \sum_{i}\bm d_{i} / N_e $.
The center-of-mass term is $F_{\mathrm{c.m.}}(z) \equiv \sigma(z)^2$, and $\sigma(z)$ is a modified Weierstrass sigma function \cite{haldane-jmp-2018}:
\begin{align}
    \sigma(z)
    &=
        \frac{\vartheta_{1} (\kappa z; \tau)}{\kappa \vartheta_1'(0; \tau)}
        \exp \left[
            i \frac{(\kappa z)^2}{\pi (\tau - \tau^{*})}
        \right].
\end{align}
Here, $\vartheta_1$ is a Jacobi theta function, $\kappa=\pi/L_1$, $L \equiv (L_x+iL_y)/\sqrt{2}\ell_B$ is the linear complex dimension of the system, with $L_1^*L_2-L_2^*L_1=2\pi i N_\phi$, and $\tau \equiv L_2/L_1$ is the modular parameter of the torus
\footnote{Since we are interested in the expectation value of Coulomb interaction which is only a function of relative coordinates and not of center of mass, we keep only the relative part of $F_{\text{CF}}$.}.
For the present calculations, we have chosen a square torus
\footnote{The particular form of the modified Weierstrass sigma-function used here is only valid for square and hexagonal tori, for more general cases see Ref.~\cite{haldane-jmp-2018}.}.

To calculate the expectation value of the Coulomb interaction in different Landau levels (ignoring Landau-level mixing), we use a Landau-level-specific Hamiltonian for $\nu=1/2$ for $n=0$, $5/2$ for $n=1$, and $9/2$ for $n=2$:
\begin{align}
\label{eqn:coulombhamiltonian}
  \mathcal H &=
    \sum_{\bm q}
    \sum_{i < j} 
      e^{i \bm q \cdot (\bm R_{i} - \bm R_{j}) }
      \widetilde{V}(\bm q)
      L_{n}^2 \left( \frac{q^2}{2} \right)e^{-q^2/2}
\end{align}
where $L_{n}(x)$ is the Laguerre polynomial of order $n$ \cite{haldane-bookchapter-1990}, and $\widetilde{V}(\bm q)=1/q$, with $n=0$ for $\nu=1/2$, $n=1$ for $\nu=5/2$, and $n=2$ for $\nu=9/2$.
However, the Monte Carlo calculation of the variational energy of this state for high LLs becomes very noisy and must be regularized.
The root of this ultraviolet behavior can be traced to strong short-range repulsions that are
generated by the Laguerre polynomials
\footnote{For example, using \textsc{mathematica} we can Fourier transform  Eq.~\eqref{eqn:coulombhamiltonian} (we drop the Gaussian and replace guiding center coordinates by ordinary ones) including the layer profile exponential factor given in the text.
It is then possible to take the thin layer limit which gives the following potential for $n=1$: $1/r+2/r^3+2.25/r^5$.}.
Fortunately, there is a physical way to regularize the Monte Carlo integration.
We introduce a short-distance cutoff $\eta$ by modifying the $1/r$ dependence of the Coulomb interactions to $1/\sqrt{r^2+\eta^2}$.
This form has been proposed to approximate the effect of finite thickness of the electron layer \cite{zhang-prb-1986}, where in magnetic length units $\eta$ is related to the average width $\bar{w}$ by $\eta=\bar{w}/2$ \cite{zhang-prb-1986}.
\begin{align}
 \widetilde{V}(\bm q) &= \frac{e^{-\eta q}}{q}.
\end{align}
The limit  $\eta\rightarrow 0$ corresponds to the pure Coulomb interaction.
We compute Fermi liquid parameters as functions of $\eta$.

We then use a variational Monte Carlo method to calculate the Fermi liquid parameters of the Coulomb interaction in  the lowest 
three half-filled Landau levels for the CF Fermi liquid state.
We follow the technique employed by \textcite{kwon-prb-1994}, which was used to study the Fermi liquid parameters of a two-dimensional electron gas.
Starting from a ``ground state'' of a Fermi sea, with $N_e=37$ filled momenta clustered around $\bm k=0$, we consider eight different low-lying quasiparticle-quasihole pair configurations labeled by $\alpha=1, \ldots, 8$ [Fig.~\ref{fig:excon}].
We then evaluate the energy (expectation value of the interacting Hamiltonian) of each of the configurations using Monte Carlo integration.
After parameterizing the energies of these configurations as a function of angle $\theta_\alpha$ between the quasiparticle and the quasihole $E_\alpha \equiv E(\theta_\alpha)$,
we fit them to the Fermi liquid energy functional
\begin{align}
 E(\theta_\alpha)
   &=
     E_0 + \epsilon_p - \epsilon_h 
     - \sum_{\ell} f_{\ell} \cos \left( \ell \theta_\alpha \right),
\end{align}
where $E_0$ is the energy of the ground state, $\epsilon_p$ and $\epsilon_h$ are the kinetic energies of the quasiparticle and the quasihole, and $f_{\ell}$ are the Fermi liquid parameters.
Since $\epsilon_p$ and $\epsilon_h$ are chosen to be equal, the angular dependence is encoded purely in $f_{\ell}$.

To test for Pomeranchuk instability, we need to normalize the Fermi liquid parameters $F_{\ell} \equiv N_F f_{\ell}$, where $N_F$ is the ``density of states'' at the Fermi energy.
Nevertheless, all other $F_\ell$'s for $\ell>1$ can be expressed in terms of $F_1$.
Unlike an ordinary Fermi liquid, however, the CF Fermi liquid wave function $\Psi_{\text{CF}}$ is explicitly $\bm K$ invariant.
By fixing $F_1=-1$ \cite{halperin-prl-1998}, we obtain the values of other Fermi liquid parameters for the composite Fermi liquid.

\begin{figure}%
\centering%
\includegraphics[width=0.35\textwidth]{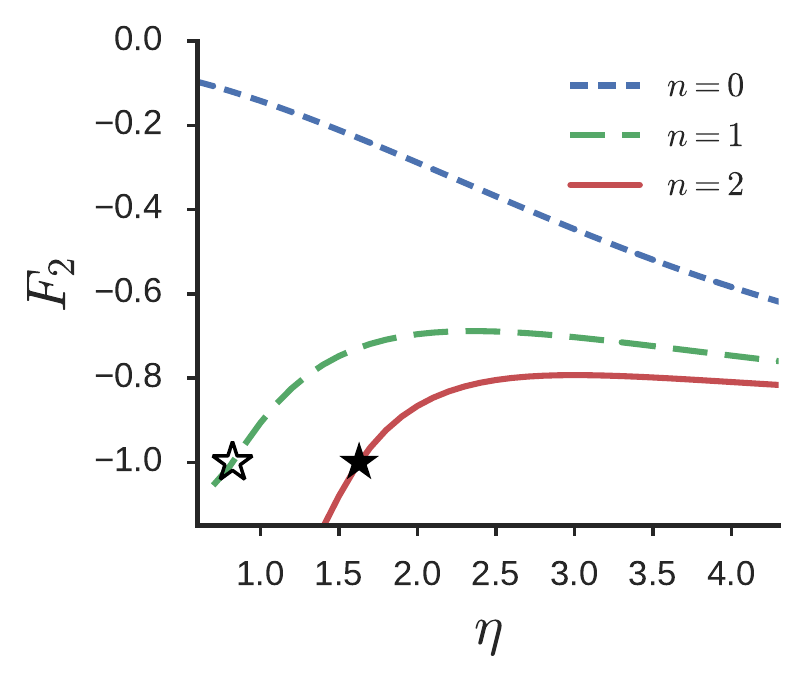}
\caption{\label{fig:flparam-all}%
$\ell=2$ Fermi liquid parameter $F_2$ for three Landau levels ($n=0, 1,$ and~$2$) plotted 
as functions of $\eta$. %
Stars mark the critical values of $\eta=\eta_c$ which yield $F_{2}=-1$
($\eta_c=0.81$ for $n=1$ and $\eta_c=1.64$ for $n=2$).
}%
\end{figure}

\begin{figure}%
\centering%
\subfigure[\label{fig:flparam-each-a}]{
\includegraphics[height=1.6in]{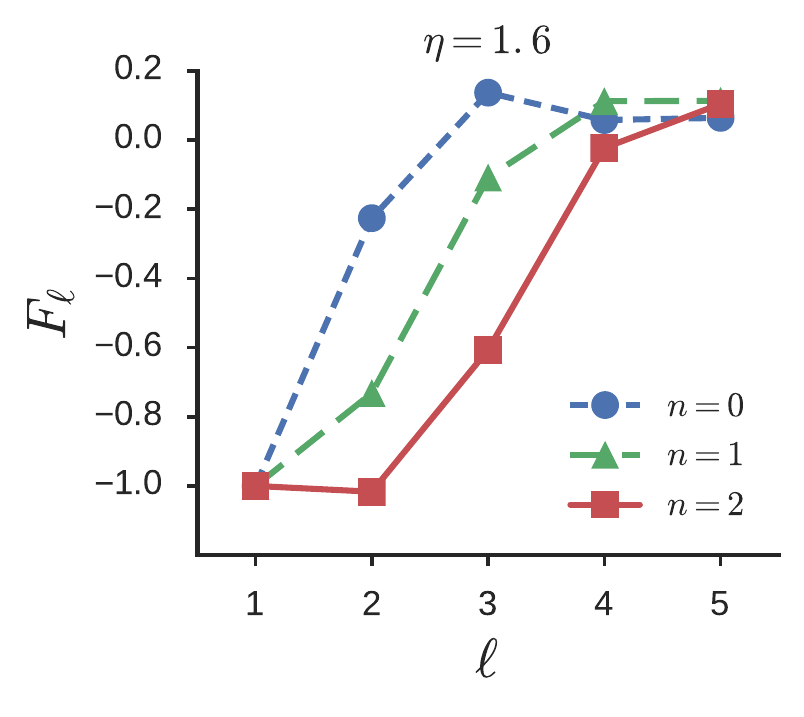}}
\subfigure[\label{fig:flparam-each-b}]{
\includegraphics[height=1.6in]{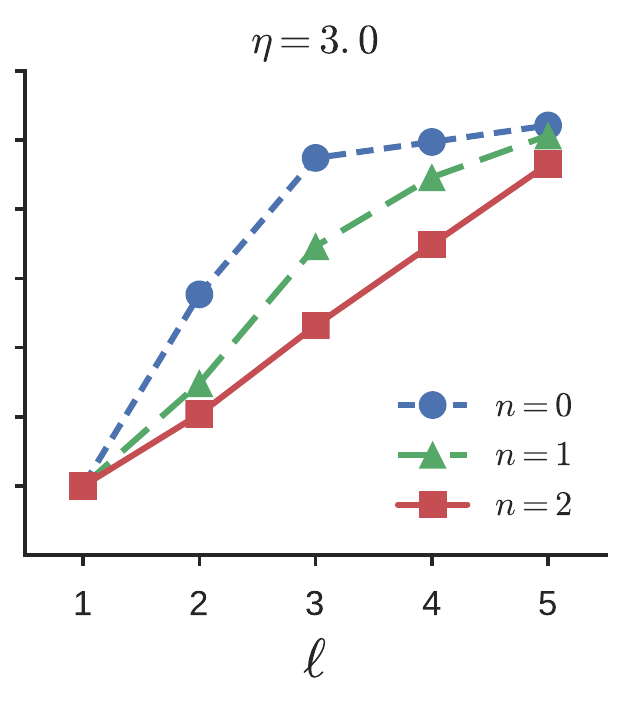}}
\caption{\label{fig:flparam-each}%
Fermi liquid parameters $F_{\ell}$ \subref{fig:flparam-each-a} slightly below the critical value of $\eta$ for $n=2$ ($\eta=1.6$) and \subref{fig:flparam-each-b} above the critical value of $\eta$ for $n=2$ ($\eta=3.0$).
For larger $\eta$, all Fermi liquid parameters are above Pomeranchuk instability point ($F_{\ell}>=-1$) and hence the system is stable against Pomeranchuk instability in any channel.
For $\eta$ slightly below the critical value for $n=2$, only the $\ell=2$ channel shows instability.
}%
\end{figure}

Our results are summarized in Fig.~\ref{fig:flparam-all}, where $F_2$ in $n=0,1,2$ Landau levels are plotted as functions of $\eta$.
The error is bound by the machine precision and the statistical error from Monte Carlo calculations is smaller than the width of the lines.

For the lowest ($n=0$) Landau level, no Pomeranchuk instability (other than $\ell=1$) is found for any value of $\eta$.
In higher ($n=1,2$) Landau levels, on the other hand, we find Pomeranchuk instability in the nematic ($\ell=2$) channel at critical values of $\eta=\eta_c$ defined by $F_2=-1$: $\eta_c=0.81$ for $n=1$ and $\eta_c=1.64$ for $n=2$.
As Fig.~\ref{fig:flparam-each} shows, Pomeranchuk instability in the nematic channel occurs over a wider range of the phenomenological cutoff parameter for $n=2$, which is consistent with the experimental observation of the QH nematic state being limited to $n=2$ under ambient pressure.
On the other hand, the fact that $n=1$ can indeed show nematic instability for $\eta < \eta_c = 0.81$ is significant in light of a recent observation \cite{samkharadze-np-2016} of transition between a fractional QH state and a nematic state at filling factor $\nu=5/2$.

The corresponding value of $\eta$ for the quantum well of width  $w_0\sim1.5\ell_B$ (or less) used by \textcite{samkharadze-np-2016} using either the model of Ref.~\cite{zhang-prb-1986} or the model of Ref.~\cite{price-prb-1984} puts the system slightly below the critical value $\eta_c$ for $n=1$.
Hence, our results taken on face value predict nematic instability even under ambient conditions for the system.
However, under these conditions, experiments as well as finite size studies \cite{rezayi-prl-2000}, unlike the $n=2$ case, do not show any sign of nematic order at $\nu=5/2$.
The gapped phase at the 5/2 Landau-level filling preempts the nematic phase.
Unfortunately, the pairing instability leading to a gapped phase at $\nu=5/2$ is inaccessible to our calculation.
Nonetheless, it is significant that we observe a Pomeranchuk instability at $\nu=5/2$, as it shows that the broken symmetry phase is, in fact, contiguous to the paired phase \cite{rezayi-prl-2000};
    under slight changes of the interaction potential, the $\ell=2$ Pomeranchuk deformation becomes a relevant perturbation.
In the \textcite{samkharadze-np-2016} experiment, hydrostatic pressure drives the  instability to the nematic phase.

We have looked into the possibility of Pomeranchuk instability in other channels.
Interestingly, we find Pomeranchuk instability only in the $\ell=2$ channel (other than $\ell=1$, which is required by the $\mathbf{K}$ invariance).
In Fig.~\ref{fig:flparam-each}, we plot $F_{\ell}$ for $\ell=1, \ldots, 5$ for $\eta=1.60<\eta_c$, and $\eta=3.00>\eta_c$, where $\eta_c$ is the critical value of the cutoff paramer  $\eta$, below which $F_2<-1$ in the $n=2$ Landau level.
For $\eta < \eta_c$ in Fig.~\ref{fig:flparam-each-a}, $F_2<-1$ for $n=2$, while all the other $F_\ell > -1$ for $\ell > 2$.
For the lowest two $n=0$ and $n=1$ levels for both values of $\eta>\eta_c$, no Pomeranchuk instability is observed.
In both cases, the Fermi liquid parameters $F_\ell$ are roughly a decreasing function of $\ell$.
In all parameter ranges we have considered, $\ell = 2$ is the leading instability with the most 
negative Fermi liquid parameter except $\ell = 1$.

The system of 37 electrons used in our calculations is sufficiently large for the purpose of detecting the Pomeranchuk instability that favors the nematic phase
\cite{[{For example, the linear dimension of our system is, in magnetic length units, about 22, which has been found to be sufficient in infinite density matrix renormalization group studies.
Specifically, these calculations use an infinite length cylinder, but limit the circumference to finite values.
The two-dimensionality of the HFLL problems are adequately captured by length scales of 20--24. See }] zaletel-prb-2015,*geraedts-s-2016}.
Our system is sufficiently large with a nearly circular Fermi surface to minimize the energy differences between particle and hole excitations \footnote{The larger size with similar properties contains 69 electrons.}.
However, for finite sizes the critical layer width would depend on the details of the Fermi surface and the geometry of torus unit cell.
Landau-level mixing, which we have ignored, will presumably also affect critical widths.
None of these effects appear to be large enough to change our main conclusions.

In summary, we explicitly calculated the Fermi liquid parameters of a CF Fermi fluid to check for the Pomeranchuk instability in a given angular momentum channel indicated by $F_{\ell}<-1$.
Ignoring Landau-level mixing, we used a Landau-level-specific Hamiltonian and took the finite quantum well thickness into account following Ref.~\cite{zhang-prb-1986}.
Our results revealed remarkable trends:
    (1) both $n=1$ and $n=2$ HFLL states exhibit nematic instability ($\ell=2$ Pomeranchuk instability) below a critical value of thickness parameters;
    (2) $n=2$ HFLL shows nematic instability at higher critical thickness leaving a wider range of thickness parameters for nematic order, whereas the $\eta_c$ for $n=1$ is below one magnetic length;
    (3) $F_{\ell}>-1$ for all $\ell>2$, ruling out all Pomeranchuk instability other than the nematic instability.
These observations are remarkably consistent with experimental observations of a nematic QH state being limited to $n=2$ HFLL under ambient conditions in that this HFLL has a much wider range of $\eta$ that shows nematic instability than that of the $n=1$ HFLL.
Also, our results predict the $\nu=1/2$ state to be stable against Pomeranchuk instability.
Our findings are qualitatively consistent with earlier observations of the QH nematic state that the anisotropic behavior is favored at smaller values of the thickness parameter \cite{doan-prb-2007,rezayi-prl-2000}.
Nevertheless our results constitute the first explicit demonstration that nematic Pomeranchuk instability can drive nematic QH states with isotropic screened Coulomb interactions.

Our finding of nematic instability in the $n=1$ Landau level for $\eta<0.8$ clearly shows that the nematic order is a contending phase for the $\nu=5/2$ state.
Recent observation of such a  transition driven by isotropic pressure \cite{samkharadze-np-2016} corroborates this picture.

However, the analysis for the case of 5/2 filling is more complicated since there could be a competition among nematic, smectic, and $p$-wave paired Moore-Read \cite{moore-npb-1991} phases.
The energy scale below which the anisotropic gapless phase has been  observed at 5/2 is more than an order of magnitude smaller than the predictions of HF approximation \cite{koulakov-prl-1996,*koulakov-prb-1996, *fogler-prb-1997}.
A similar trend appears for the 9/2 filling \cite{fradkin-arcmp-2010}.
To our knowledge, there is no method of detecting  pairing instability from Fermi liquid parameters.
The question as to  which of these phases prevails can be answered by energetic considerations, which is beyond the scope of this Letter.
Finite size calculations show that under ambient presure and untilted magnetic field, the paired phase appears to be dominant \cite{rezayi-prl-2000,papic-prb-2009,peterson-prl-2008}, in agreement with experiment.

\begin{acknowledgments}
We thank Eduardo Fradkin for suggesting this calculation.
We thank Jim Eisenstein, Steve Kivelson, and Mike Manfra for discussions.
K.L. acknowledges support from National Science Foundation Grant No.~DMR-1629382.
E.-A. K. was supported by the U.S. Department of Energy, Office of Basic Energy Sciences, Division of Materials Science and Engineering under Award No. DE-SC0010313.
E.-A. K. also acknowledges Simons Fellow in Theoretical Physics Grant No. 392182 and hospitality of the KITP supported by Grant No. NSF PHY11-25915.
F.D.M. H. and E.H. R. are supported by Department of Energy Grant No. \protect{DE-SC0002140}.
\end{acknowledgments}


%

\end{document}